\documentclass[final,5p,twocolumn,lettersize]{elsarticle}
\usepackage{graphicx}
\biboptions{sort&compress}

\usepackage{amssymb}
\newcommand{\cobaltite}{Co$_3$O$_4$}
\newcommand{\spinel}{MgAl$_2$O$_4$}

\journal{Surface Science}

\begin{document}

\begin{frontmatter}



\title{Interface and electronic characterization of thin epitaxial {C}o$_3${O}$_4$ films}

\author[addr1,addr0]{C.~A.~F.~Vaz\corref{cor1}} \cortext[cor1]{Corresponding author. Email: carlos.vaz@yale.edu}%
\author[addr1,addr0]{H.-Q.~Wang}
\author[addr1,addr0]{C.~H.~Ahn}
\author[addr1,addr0]{V.~E.~Henrich}
\author[addr2,addr0]{M.~Z.~Baykara}
\author[addr2,addr0]{T.~C.~Schwendemann}
\author[addr2,addr0]{N.~Pilet}
\author[addr2,addr0]{B.~J.~Albers}
\author[addr2,addr0]{U.~D.~Schwarz}
\author[addr3,addr0]{L.~H.~Zhang}
\author[addr3,addr0]{Y.~Zhu}
\author[addr4,addr0]{J.~Wang}
\author[addr5,addr0]{E.~I.~Altman}

\address[addr1]{Department of Applied Physics, Yale University, New Haven, Connecticut 06520}%
\address[addr2]{Department of Mechanical Engineering, Yale University, New Haven, Connecticut 06520}%
\address[addr3]{Brookhaven National Laboratory, Building 480, Upton, New York 11973}%
\address[addr4]{Department of Physics, Yale University, New Haven, Connecticut 06520}%
\address[addr5]{Department of Chemical Engineering, Yale University, New Haven, Connecticut 06520}%
\address[addr0]{Center for Research on Interface Structures and Phenomena (CRISP), Yale University, New Haven, Connecticut 06520}%

\begin{abstract}
The interface and electronic structure of thin ($\sim$20-74 nm)
\cobaltite(110) epitaxial films grown by oxygen-assisted molecular
beam epitaxy on \spinel(110) single crystal substrates have been
investigated by means of real and reciprocal space techniques.
As-grown film surfaces are found to be relatively disordered and
exhibit an oblique low energy electron diffraction (LEED) pattern
associated with the O-rich CoO$_2$ bulk termination of the (110)
surface. Interface and bulk film structure are found to improve
significantly with post-growth annealing at 820 K in air and display
sharp rectangular LEED patterns, suggesting a surface stoichiometry
of the alternative Co$_2$O$_2$ bulk termination of the (110)
surface. Non-contact atomic force microscopy demonstrates the
presence of wide terraces separated by atomic steps in the annealed
films that are not present in the as-grown structures; the step
height of $\approx 2.7$ \AA\ corresponds to two atomic layers and
confirms a single termination for the annealed films, consistent
with the LEED results. A model of the ($1\times 1$) surfaces that
allows for compensation of the polar surfaces is presented.
\end{abstract}


\end{frontmatter}


% main text
\section{Introduction}
%\label{}

The oxides of the 3d transition metals form an important class of
materials with properties that depend sensitively on the cationic
oxidation state and the electronic environment. As a consequence,
these compounds display a  multiplicity of magnetic, electronic and
catalytic behavior that makes them interesting from both fundamental
and practical perspectives. In particular, the ongoing trend towards
controlling the electronic properties of materials at the nanoscale
implies that, in addition to the development of methods for the
fabrication and growth of high quality thin films, an understanding
of the physical mechanisms underlying the properties of such systems
at the atomic scale are key. In this paper we show that the surface
and bulk properties of [110]-oriented \cobaltite\ thin films depend
sensitively on growth conditions and post-growth annealing. In
particular, we show that the surface termination can be switched
between the two possible bulk-terminations of [110]-oriented
spinels.

Cobalt, like most 3d transition metal elements, can exist in more
than one oxidation state. Of the two stable cobalt oxides, the mixed
valence compound, Co$^{2+}$Co$_2^{3+}$O$_4$, is stable at ambient
pressure and temperature and crystallizes in the spinel structure.
Meanwhile, the high temperature CoO phase crystallizes in the rock
salt structure. Both oxides are antiferromagnetic at low
temperatures, with N\'eel temperatures of approximately 40 and 290
K, respectively
\cite{Bizette46,Bizette51,Blanchetais51,Roth58,SSW51,Roth64,SH76,JRB+01}.

Surfaces and interfaces of \cobaltite\ are complicated by the fact
that all the low index planes of the spinel structure are polar.
Therefore, the clean, bulk-terminated crystal surfaces have
divergent electrostatic surface energies due to a
thickness-dependent electric dipole of the crystal
\cite{Tasker79,Noguera00,GFN08}. Charge compensation mechanisms that
lead to a finite dipole may result in important modifications of the
surface geometric and electronic structure, including changes in the
valence state of surface ions, surface reconstructions, surface
roughening and faceting, among others
\cite{HKWG78,Tasker79,Noguera00,JPS+02,LPP+05,GFN08}. One topical
example of an electronic modification is the recent observation of
metallic interface states in SrTiO$_3$/LaAlO$_3$ heterostructures,
believed to originate from the polar discontinuity across the
interface \cite{SS08}. There is therefore general interest in
studying the surface and interface structure of polar oxides.

\begin{figure}[h]
\begin{centering}
\includegraphics*[width=\columnwidth]{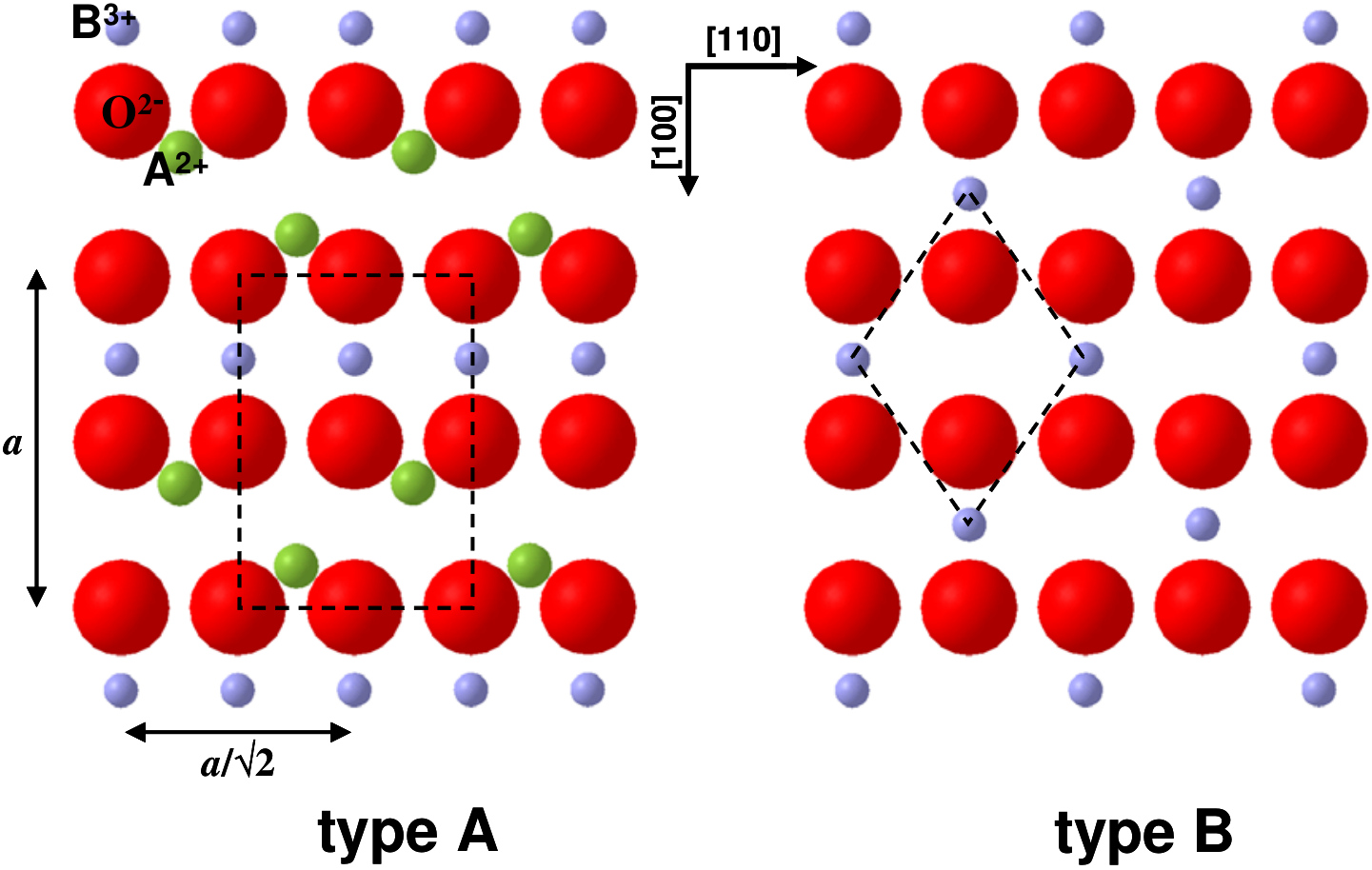}
\caption{(Color online) The two possible bulk crystal terminations
of the spinel structure A$^{2+}$B$_2^{3+}$O$_4$ along the (110)
plane, where the A cations occupy tetrahedral positions and the B
cations occupy octahedral positions; the dashed lines correspond to
the surface primitive unit cells. In type A, both 2+ and 3+ cations
are present, while type B has only 3+ cations. Both surfaces are
polar.} \label{fig:spinel_surface}
\end{centering}
\end{figure}

Here, we consider the surface and interface properties of
[110]-oriented epitaxial films of the prototypical \cobaltite\
spinel grown on \spinel(110) substrates.  While the growth of
polycrystalline \cobaltite\ films has been reported extensively
\cite{PS81,YTH03,BGS+05,RLH06}, the growth of epitaxial \cobaltite\
films has been studied much less often. In one instance, epitaxial
\cobaltite\ films up to 5 \AA\ thick have been grown on CoO(001)
single crystals by oxidation at high temperatures in an oxygen
atmosphere \cite{CNL96,LAC+99}. Atomic layer deposition has also
been used to grow epitaxial \cobaltite\ films on MgO(001)
\cite{RLH06}. However, the use of \spinel\ substrates has several
unique advantages; both materials are normal spinels and have
lattice constants that match each other almost exactly, $a=8.086$
\AA\ for \cobaltite\ \cite{PBBC80} and $a=8.0858$ \AA\ for \spinel\
\cite{IYY+00,DGS+04}. The same crystal structure should also
preclude the formation of antiphase boundaries that originate when
lower symmetry structures are grown on higher symmetry surfaces, and
the good chemical and thermal stability of \spinel\ guarantees no
interdiffusion at high growth temperatures. The (110) surface is
also characterized by having a four-repeat period as opposed to the
eight-period repeat of the (100) surface, and therefore should be
less susceptible to stacking faults and antiphase boundary
formation, leading in principle to fewer defects in the film. As
pictured in Fig.~\ref{fig:spinel_surface}, in the [110]-direction
the spinel structure is composed of type A planes with a
Co$_2^{2+}$Co$_2^{3+}$O$_4$ stoichiometry and a formal charge of +2
per surface unit cell alternating with Co$_2^{3+}$O$_4$ type B
planes with a formal charge of --2. The orientation is therefore
polar, and faceting or reconstruction of the film surfaces might be
expected.

In this paper we show that \cobaltite(110) thin films can be grown
epitaxially on \spinel(110) substrates by oxygen assisted molecular
beam epitaxy.  Despite the expectation that the polar surfaces would
reconstruct, we observe only ($1\times 1$) surface diffraction
patterns with no evidence of periodic reconstructions. Although the
as-grown films display rough surfaces and bulk defects, these
features can be largely eliminated by post-growth annealing in air,
which leads to well ordered, atomically flat surfaces and
interfaces. Such well defined surfaces have allowed the growth of
stoichiometric epitaxial PdO thin films for surface reaction studies
\cite{VYH+08}. Interestingly, we find that annealing appears to
change the surface termination from a ($1\times 1$) B-type
termination to a ($1\times 1$) A-type termination.  A model of the
($1\times 1$) surfaces that allows for compensation of the polar
surfaces will be presented.

\section{Sample growth and characterization}

\begin{figure}[t!bh]
\begin{centering}
\includegraphics*[width=\columnwidth]{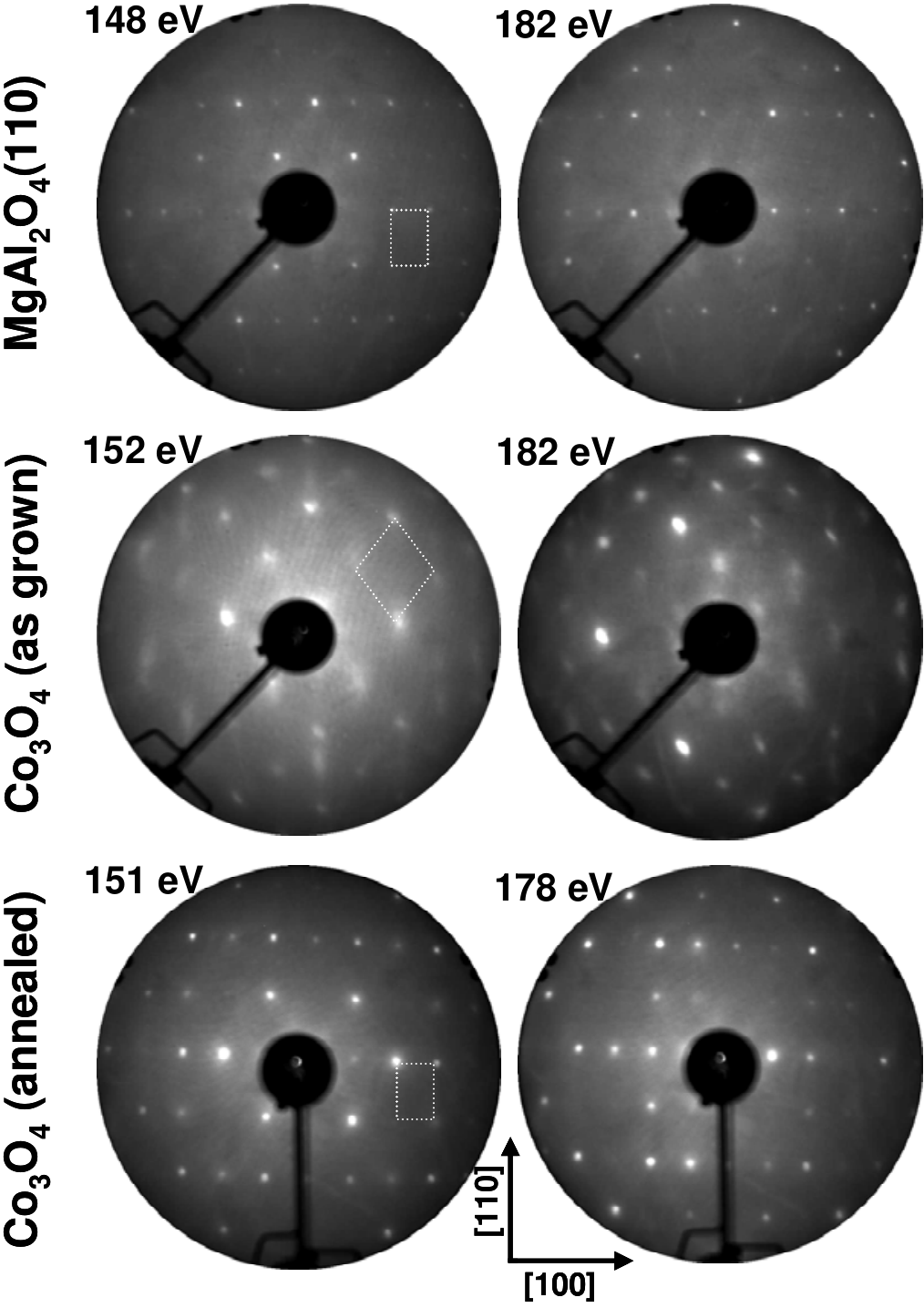}
\caption{Low energy electron diffraction (LEED) patterns of the
\spinel(110) and \cobaltite\ films before and after annealing, at
two selected electron beam energies. The crystal orientation,
inferred from the LEED pattern, is shown in the inset. Unit cells of
the LEED pattern are drawn in white. For the \spinel\ and annealed
\cobaltite\ LEED patterns, the dimensions of the unit cell (in
reciprocal space) correspond to the unit cell of the surface,
$a\times a/\sqrt{2}$, as estimated from the scattering geometry of
our LEED system.} \label{fig:LEED}
\end{centering}
\end{figure}

\begin{figure*}[t!bh]
\begin{centering}
\includegraphics*[width=15cm]{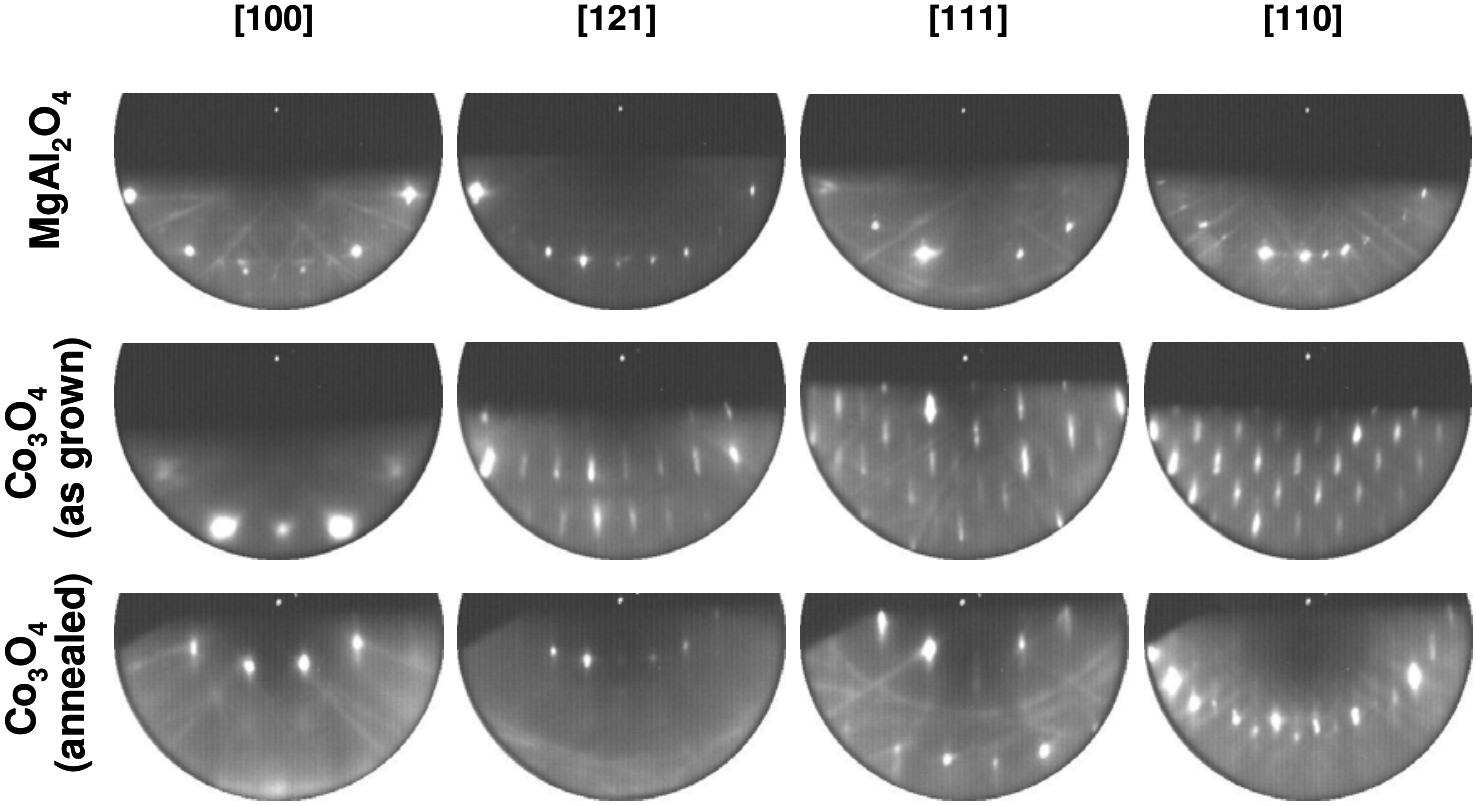}
\caption{Reflection high energy electron diffraction (RHEED)
patterns of the \spinel(110) substrate and \cobaltite\ films before
and after annealing, along different azimuths (parallel to the
electron beam, set at a grazing angle of incidence). The angle of
incidence changes slightly with azimuth due to eccentricity in the
sample holder.} \label{fig:RHEED}
\end{centering}
\end{figure*}

\spinel(110) single crystals were used as substrates for the growth
of \cobaltite\ due to the small lattice mismatch of --0.05\% and
good thermal and chemical stability. The substrates were first
outgassed and cleaned {\it in situ} with an O-plasma at 770 K for 30
min. Such treatment renders the \spinel\ surface free of C
contaminants, as determined by Auger electron spectroscopy (AES);
the only impurities detected consist of 2-3 at.~\% Ca to within the
probing depth of AES, about 3 nm. The high quality of the
\spinel(110) surface crystallinity was confirmed by low energy
electron diffraction (LEED) and reflection high energy electron
diffraction (RHEED), which display patterns characteristic of highly
ordered surfaces (see Figs.~\ref{fig:LEED} and \ref{fig:RHEED}). The
\cobaltite\ films were grown by oxygen assisted molecular beam
epitaxy by simultaneous exposure of the substrate to a thermally
evaporated Co atomic beam and an atomic O flux. Initial growth
studies indicated that better films were obtained when growth was
started at 770 K and then lowered to 570 K after about 1 nm
deposition, as judged by RHEED. The initial higher temperature
promotes surface diffusion, while continued growth at this
temperature caused CoO to form. The oxygen partial pressure during
growth was $3\times 10^{-5}$ mbar, and the electron cyclotron
resonance oxygen plasma source magnetron was run at 125 W, yielding
an atomic O flux of the order of $1\times 10^{14}$ cm$^{-2}$s$^{-1}$
at the sample \cite{GKA05,AWN+00}. The Co evaporation rate was
typically 2 \AA/min, and film thicknesses were estimated by means of
a calibrated quartz crystal balance. Sample growth was monitored
with RHEED, and after layer completion, the \cobaltite\ film
crystallinity and electronic structure were determined by RHEED,
LEED, x-ray photoemission (XPS), and Auger electron spectroscopy
(AES). AES indicates that no significant amounts of impurities were
present (none were detected within the accuracy of our spectrometer,
equipped with a double-pass cylinder mirror analyzer).

After film growth, the samples were characterized using a variety of
structural techniques, before and after annealing, including x-ray
diffraction and reflectometry, atomic force microscopy and
transmission electron microscopy. Sample annealing was performed at
820 K for 14 h in flowing air; this temperature and oxygen partial
pressure favor the formation of \cobaltite\ over CoO
\cite{TK55,KY81a,OS92a}. For this study, eight sets of samples were
grown and characterized, and the results were found to be consistent
and reproducible for all samples, in particular the LEED, RHEED,
XPS, atomic force microscopy (AFM) and AES data. Non-contact AFM
data were taken on two different samples, and transmission electron
microscopy (TEM) measurements were made on one sample only (for TEM,
one half was annealed after cleaving, for a comparative study). In
the following, the LEED and RHEED data correspond to the same
sample, for better comparison of the electron diffraction data.

Typical LEED and RHEED patterns of the \cobaltite\ films after
growth are shown in Figs.~\ref{fig:LEED} and \ref{fig:RHEED},
respectively. Compared with the \spinel\ LEED patterns, the
diffraction spots of the as-grown \cobaltite\ film are much broader,
and the background is also more intense, indicating that the
as-grown \cobaltite\ films have a significant amount of surface
disorder (severe charging of the surface also contributes to the
poor patterns, especially at lower electron beam energies). The
RHEED patterns exhibit streaky and relatively broad diffraction
spots, characteristic of a three dimensional growth mode and a
relatively disordered surface. However, Kikuchi lines are observed
along some directions (the observation of Kikuchi patterns in RHEED
has been associated with the presence of relatively smooth surfaces
\cite{Braun99}). Both RHEED and LEED data indicate that, despite the
local surface disorder, long range order is preserved.

The annealing process induces significant transformations in the
film structure, as indicated by different and much sharper LEED and
RHEED patterns. In fact, the quality of these patterns becomes
comparable to that of the \spinel\ single crystal substrate surface.
For the annealed \cobaltite\ surface, the LEED background intensity
is very low, with very sharp diffraction spots, and our data show
that in the range from 40 eV to 150 eV there are no changes in the
LEED pattern and no discernible increase in background intensity.
One striking feature of the LEED patterns is the different symmetry
exhibited by the as-grown and annealed films: while the former
exhibit an oblique unit cell, the latter is found to exhibit
predominantly a rectangular unit cell. Such LEED patterns are
compatible with the two possible terminations of the (110) plane of
the spinel structure: type A termination for the annealed films,
exhibiting a rectangular unit cell, and type B for the as-grown
films, with an oblique unit cell (see
Fig.~\ref{fig:spinel_surface}).

\begin{figure}[t!bh]
\begin{centering}
\includegraphics*[width=\columnwidth]{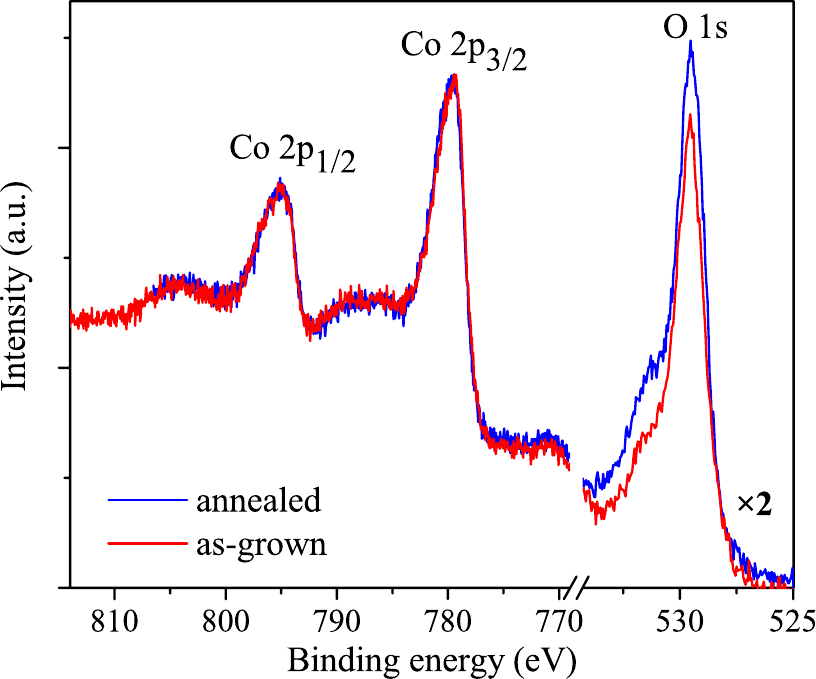}
\caption{(Color online) Core level x-ray photoelectron spectra for
as-grown and annealed \cobaltite\ films.} \label{fig:XPS}
\end{centering}
\end{figure}

X-ray photoemission spectroscopy measurements of the \cobaltite\
films after growth and after annealing were carried out to assess
the film stoichiometry. The XPS spectra were obtained using the Mg
K$_\alpha$ line ($h\nu = 1253.6$ eV) of a double anode x-ray source
and a double pass cylinder mirror analyzer ($\Phi$ 15-255G) set at a
pass energy of 25 eV (energy resolution of about 0.8 eV). High
resolution XPS spectra of the O 1s and Co 2p lines of the as-grown
and annealed films are shown in Fig.~\ref{fig:XPS}. Corrections to
the data include a five-point adjacent smoothing, satellite
correction and correction of energy shifts due to charging (aligned
with respect to the Co 2p peaks, using the energy assignments given
in \cite{CBR76,HU77}). One observation is that the Co 2p spectra for
both samples are identical, showing that no significant changes in
stoichiometry or in the ionic state of the Co cations occur as a
consequence of annealing. A second observation is that the Co 2p
spectra are characteristic of a \cobaltite\ ionic environment
\cite{CBR76,HU77}, with strongly suppressed shake-up peaks compared
to those of CoO
\cite{BGD75,CBR76,HU77,OS92a,LAC+99,WHMS04,WAH08,PMCL08}. The O 1s
photoemission line is also similar before and after annealing,
although some variation in the shape of the O 1s line is observed.
This could be related, in some instances, to the effect of the
O-plasma cleaning procedure (at room temperature) before XPS spectra
collection: given the surface sensitivity of XPS, small variations
in the amount of surface adsorbed O may lead to slight variations in
the O 1s line; in fact, the additional peak observed at higher
binding energies has been attributed to adsorbed oxygen
\cite{CBR76,JD79,KBR+84,KGBW93,GKBW95,JFEG95,CNL96,PMCL08}, although
surface hydroxylation, which also yields a similar peak
\cite{HU77,PMCL08}, cannot be ruled out.

\begin{figure}[t!bh]
\begin{centering}
\includegraphics*[width=\columnwidth]{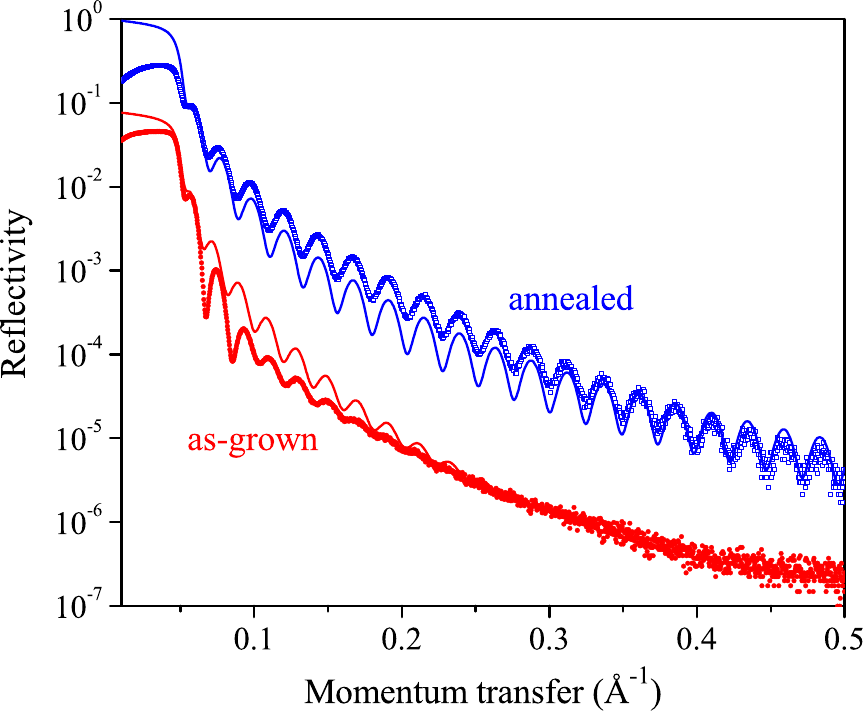}
\caption{(Color online) x-ray reflectivity spectra (symbols) for the
as-grown and annealed \cobaltite\ films. Data have been shifted by a
factor of 10 for convenient data display; solid lines are fits to
the data and give a thickness of 292 \AA\ for the as-grown film and
255 \AA\ for the annealed sample.} \label{fig:XRR}
\end{centering}
\end{figure}

Structural characterization of the \cobaltite\ films was carried out
{\it ex situ} by x-ray scattering measurements on a Shidmazu
diffractometer using the Cu K$_\alpha$ line
($\lambda_{\mathrm{K}\alpha 1} = 1.540606$ \AA) with a Ni filter to
remove the sharp Cu K$_\beta$ lines. x-ray diffraction (XRD) spectra
of the as-grown and annealed films show only diffraction peaks
associated with the (110) planes of \spinel; this also indicates
that the cobalt oxide grown is \cobaltite, since only this phase has
diffraction peaks that coincide with those of \spinel. The spectra
for the as-grown and annealed films are identical, although the
diffraction lines of the \cobaltite\ and \spinel\ cannot be
discriminated due to the closeness of the lattice constants of the
two structures. Rocking curves of the (220) and (440) peaks show a
double peak in the as-grown sample, which is attributed to the
presence of twin domains in the \spinel\ substrate; this double peak
feature is absent in the annealed sample, indicating that the
\spinel\ domains have merged upon annealing. The full width at half
maximum (FWHM) of these peaks is below the instrumental resolution
of the diffractometer, 0.02$^\mathrm{o}$. More significant are the
x-ray reflectivity (XRR) spectra for the as-grown and annealed
films. In this scattering geometry, interference between the film
interfaces gives rise to oscillations in the reflected intensity
(Kiessig fringes), whose amplitude depends on the difference in
potential between interfaces; interface roughness dampens these
oscillations and affects the rate of decay in intensity with
momentum transfer. Below the critical angle for total reflection,
all light is reflected off the surface. The XRR spectra of as-grown
and annealed \cobaltite\ films are shown in Fig.~\ref{fig:XRR}. One
finds that the reflected intensity of the as-grown film falls off
rapidly, with a strong dampening in the fringe oscillations,
suggesting relatively rough interfaces; the annealed film, however,
shows undamped oscillations over the entire momentum transfer range
probed, up to 0.5 \AA$^{-1}$. Fits to the data were performed using
bulk scattering factors and, starting with nominal thicknesses and
initial guesses for the interface roughness parameters, by modifying
the initial parameters until a good visual fit to the data was
obtained; the accuracy of the fits are within $\pm 5$ \AA\ for the
interface roughness and $\pm 2$ \AA\ for the thickness (which does
not take into account systematic errors, such as sample
displacement, which occurs if the sample is not in the center of the
goniometer circle). The fits to the experimental data suggest that
the interface roughness has decreased from 10-20 \AA\ for the
as-grown film to 0-2 \AA\ after annealing; in addition, it is found
that both surface and interface interface roughness parameters have
changed, indicating that structural changes occur throughout the
whole film following the annealing procedure.

\begin{figure}[t!bh]
\begin{centering}
\includegraphics*[width=0.49\columnwidth]{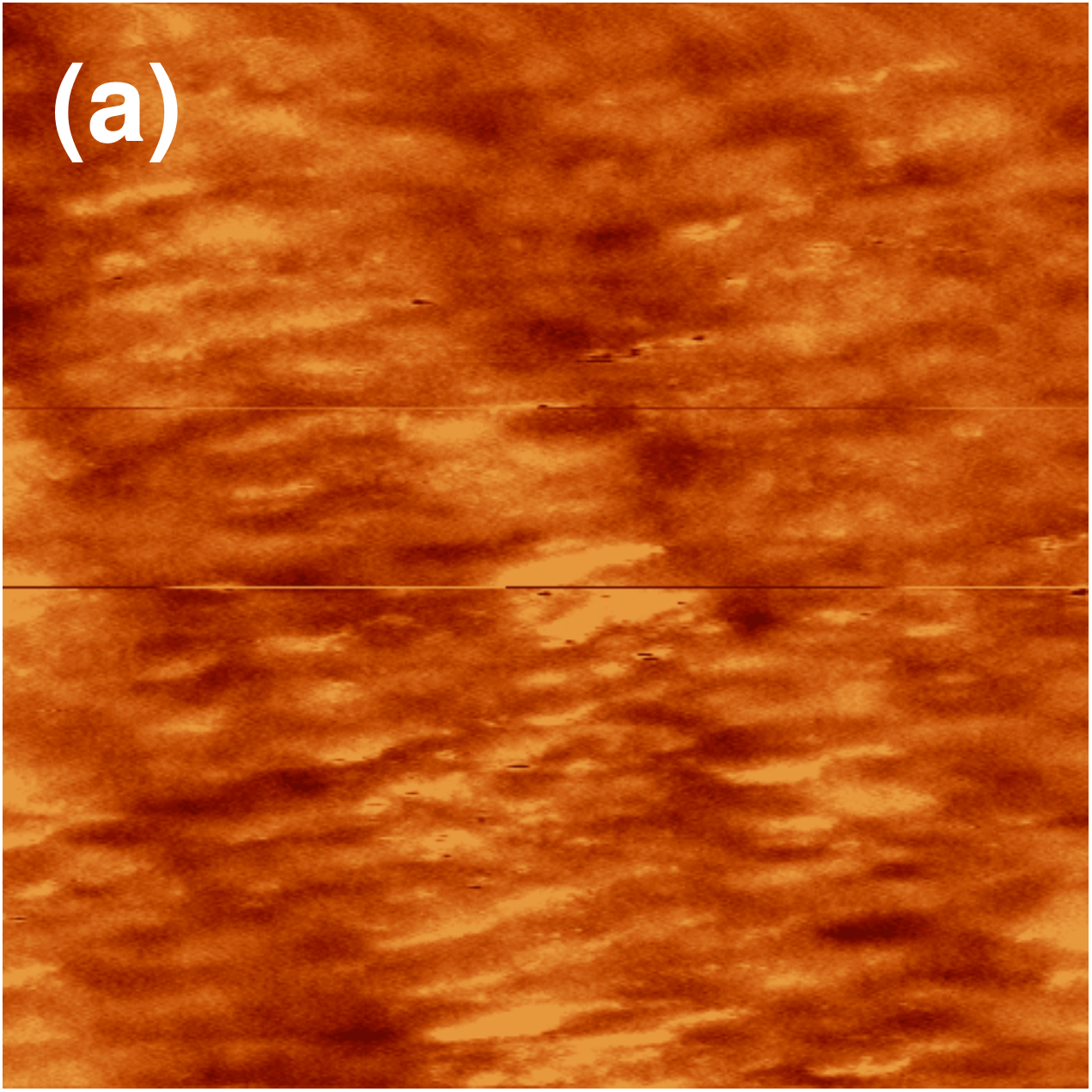}
\includegraphics*[width=0.49\columnwidth]{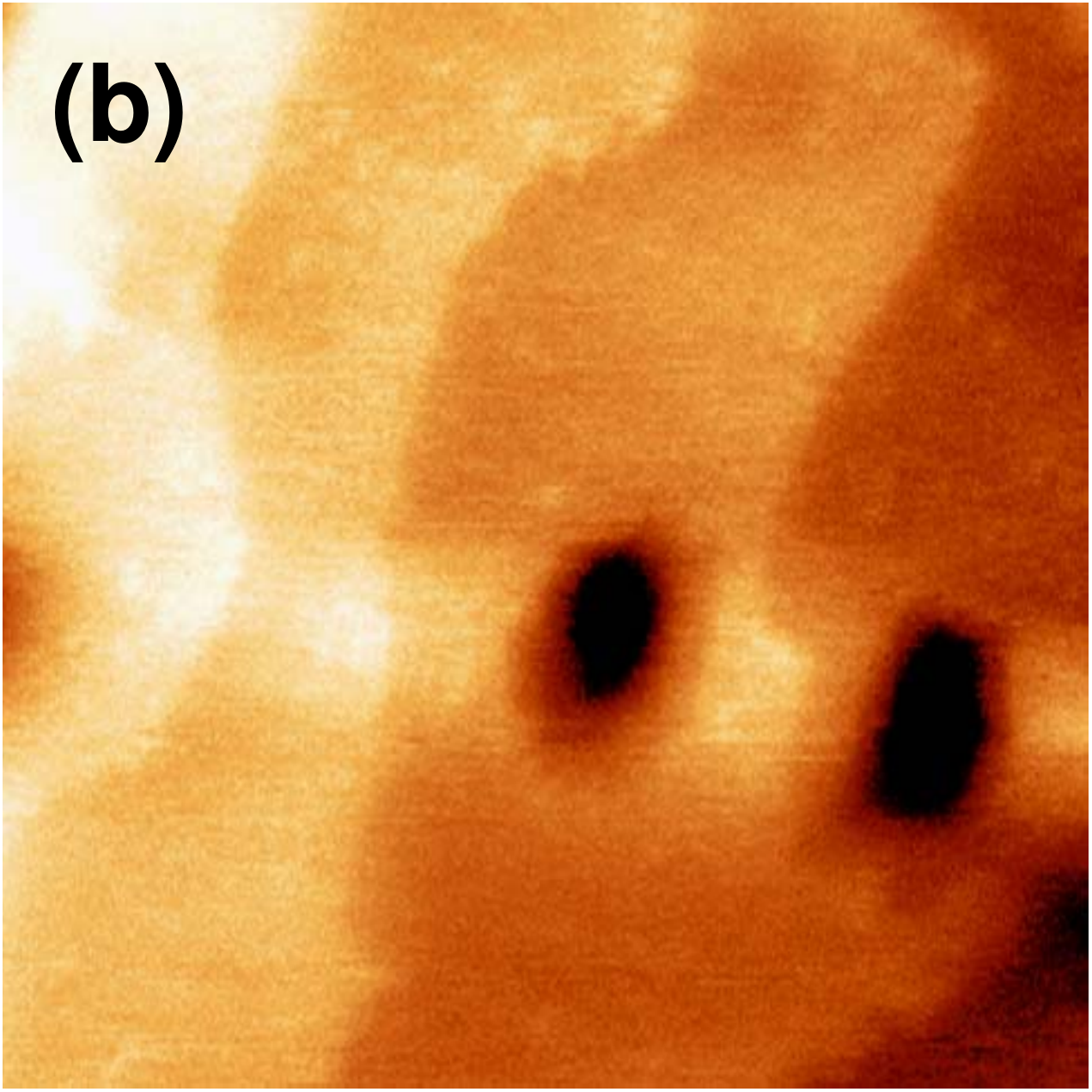}\\
\hspace*{0.45\columnwidth}
\includegraphics*[width=0.54\columnwidth]{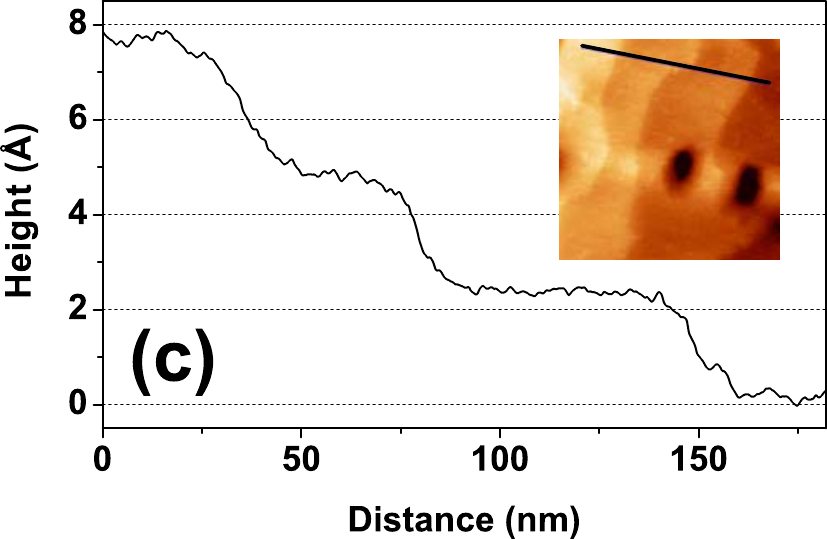}
%\includegraphics*[width=4.25cm]{asgrownz8Astp_Fig6a.eps}
%\includegraphics*[width=4.25cm]{annealingz8Astp_Fig6b.eps}\\
%\hspace*{3.86cm}
%\includegraphics*[width=4.64cm]{NCAFM_profile_Fig6c.eps}
\caption{(Color online) (a) High resolution non-contact atomic force
microscopy image of the as-grown \cobaltite\ film acquired under
ultrahigh vacuum conditions at 77 K, recorded at $\Delta f = -0.95$
Hz with a tuning fork resonance frequency of 29805 Hz, scan speed of
75 nm/s, and oscillation amplitude of 1.30 nm; the image size is
$300 \times 300$ nm$^2$. (b) NC-AFM image for the annealed
\cobaltite\ film acquired under ultrahigh vacuum conditions at 6 K,
recorded at $\Delta f = -2.6$ Hz, tuning fork resonance frequency of
30001 Hz, scan speed of 200 nm/s, and oscillation amplitude of 0.25
nm; the image size is $200 \times 200$ nm$^2$. The vertical scale
range is from 0-8 \AA\ in both cases. (c) Step profile of the
terraces observed in (b) for the annealed \cobaltite\ surface.}
\label{fig:NC-AFM}
\end{centering}
\end{figure}

The surface morphology was also probed using real space techniques,
including atomic force microscopy (AFM) at room temperature and high
resolution non-contact atomic force microscopy (NC-AFM) at low
temperatures. AFM in contact mode was used to obtain a large area
scan of the surface morphology; these scans show that annealing
induces a significant surface smoothing, with the average roughness
(measured over a 4 $\mu$m$^2$ area) decreasing from 2.5 \AA\ to 1.5
\AA\ upon annealing. Nanometer scale topographical measurements were
performed on the as-grown and annealed \cobaltite\ thin films using
a home built, low-temperature, ultrahigh vacuum atomic force
microscope, described in detail elsewhere \cite{ALS+08}. The
microscope operates in non-contact mode \cite{MWM02} using the
frequency modulation technique \cite{AGHR91} with a quartz tuning
fork and an electrochemically etched Pt/Ir tip force sensor
\cite{Giessibl00,GHH+04}. Prior to the measurements, the sample was
cleaned  in the instrument preparation chamber under an oxygen
plasma (at $5\times 10^{-7}$ mbar pressure) at $\sim$370 K for 45
min for the as-grown sample and 3 h for the annealed sample. The
results of both measurements are shown in Fig.~\ref{fig:NC-AFM} and
provide striking evidence for the improvement in surface morphology
upon annealing. The surface topography of the as-grown film is
irregular, with elongated features set preferentially along one
direction, which the RHEED data suggests to be along the in-plane
[100] direction. In contrast, the annealed surface exhibits well
defined atomic steps, confirming the atomic smoothness suggested by
the reciprocal space techniques. Figure~\ref{fig:NC-AFM}(c) shows a
line profile over four terraces, spanning about 0.8 nm in height,
giving a step height of $\sim$0.27 nm, which corresponds
approximately to two \cobaltite(110) atomic planes, 0.286 nm in
height. This result confirms a single termination for the annealed
films, consistent with the LEED results.

\begin{figure*}[t!bh]
\begin{centering}
\includegraphics*[width=14cm]{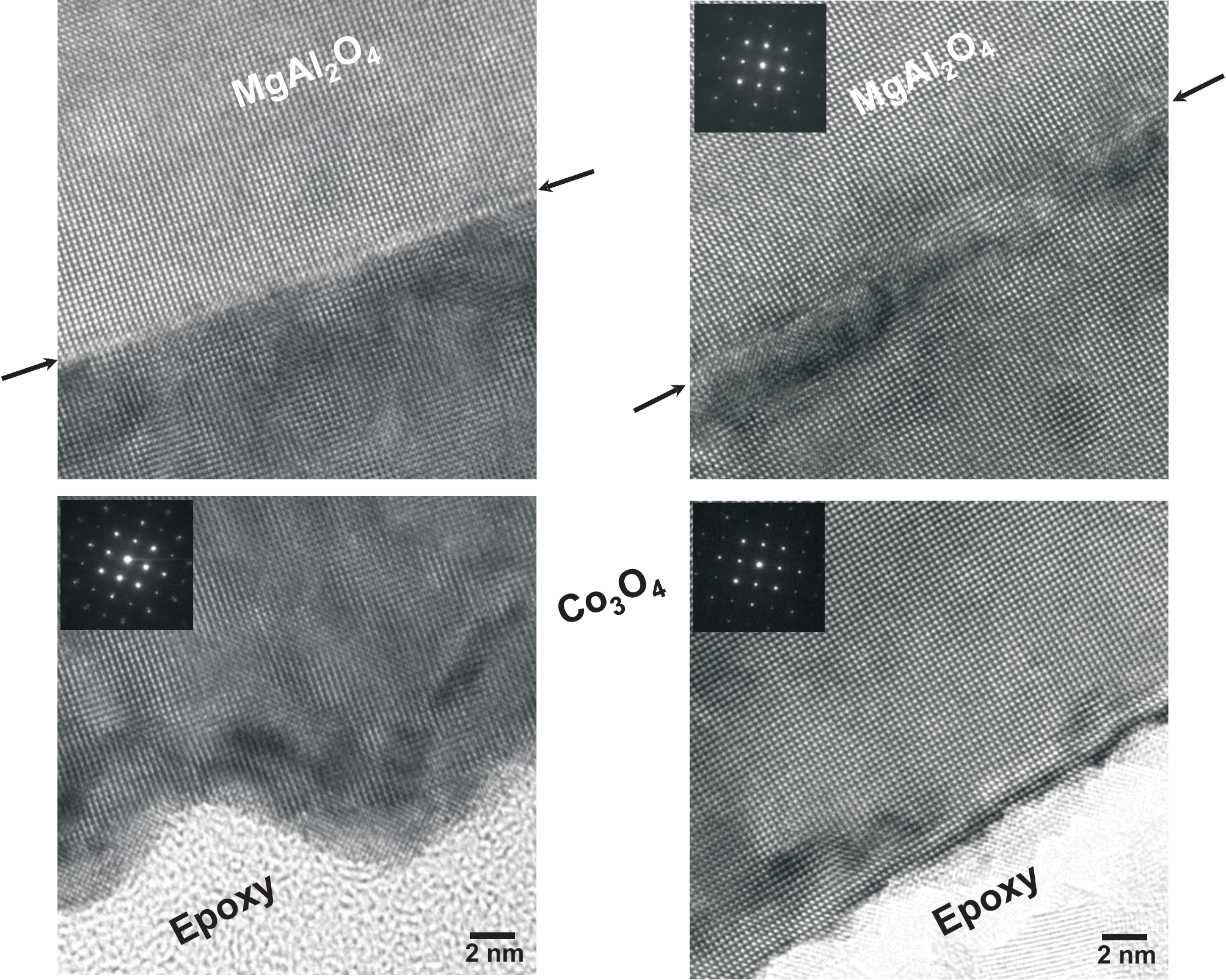}
\caption{Cross-sectional transmission electron micrographs of the
as-grown (left) and annealed (right) 25 nm \cobaltite\ films along
the (100) plane at the \spinel/\cobaltite\ interface (top) and
across the \cobaltite/epoxy interface (bottom). The white dot
pattern distance along the [110] direction (along the direction of
the arrows) corresponds to the oxygen sublattice distance along the
[110] direction, $\sqrt{2}a^\mathrm{MgAl_2O_4}/4 = 2.86$ \AA. The
arrows on the top images indicate the approximate position of the
interface. Insets correspond to electron diffraction patterns of the
\spinel\ substrate (top image) and of the \cobaltite\ and \spinel\
(bottom images).} \label{fig:TEM}
\end{centering}
\end{figure*}

In addition to the surface characterization, we have carried out
cross-sectional high resolution transmission electron microscopy
(TEM) measurements in order to obtain a more detailed knowledge of
the crystalline structure at the film interface and in the bulk of
the film. Sample preparation involved slicing, polishing, dimpling
and Ar-ion milling in a liquid nitrogen cooled sample stage. The TEM
images were obtained in a JEOL 3000F microscope at 300 kV with a
point resolution of 0.165 nm. Representative TEM images obtained for
the as-grown and annealed samples are shown in Fig.~\ref{fig:TEM},
where several modifications in the film structure upon annealing are
evident, both at the interface and in the bulk of the films: (i) one
observes that while the \cobaltite/\spinel\ interfaces are well
defined for both the as-grown and annealed films, in the regions of
2-3 unit cells from the interfaces the details of the interfacial
structure differ significantly in that the atomic positions, as
represented by the white dots, seem off registry compared with the
regions away from the interface for the as-grown film, suggesting
the presence of interfacial strain, while little registry deviation
is observed for the annealed film. (ii) In the bulk of the as-grown
film, the TEM images show speckle contrast, indicating the presence
of lattice distortion as well as variation in sectional thickness,
which could also indicate the presence of structural disorder along
the in-plane [100] direction, as suggested by the observation of the
broadening of the Bragg reflections seen in the inset to
Fig.~\ref{fig:TEM} (top left). For the annealed film, the integrity
of the crystal structure in the bulk of the film is nearly perfect.
(iii) At the surface of the film, one finds that the as-grown film
has a relatively large surface roughness; the clearly defined side
profile indicates that the roughness is oriented along the [110]
direction, as also suggested by the RHEED and NC-AFM data. The
surface atomic structure of the as-grown film seems to be strongly
modified at the top of the islands as compared to the bulk of the
film. For the annealed film, the top surface is terminated abruptly
by relatively flat plateaus, in agreement with the results of the
NC-AFM measurements. From TEM images taken at lower resolutions, we
observe that the \cobaltite\ layer thickness remains constant before
and after annealing, at 25 nm, with a roughness modulation of about
3.3 nm in  amplitude for the as-grown film (this is not in
disagreement with the AFM roughness values: more localized probes
tend to give higher roughness values due to higher spatial
resolution). One effect of annealing is that of reducing drastically
the surface roughness, which indicates a significant thermally
activated mass transport during annealing; given the small lattice
mismatch between \cobaltite\ and \spinel, the state of lowest
surface energy (area) seems to be that of thermodynamic equilibrium,
while the roughening process must be kinetically set during the film
growth. Overall, the TEM measurements indicate that post-growth
annealing results in films that have improved crystal structure and
atomically sharp and flat interfaces.

\section{Discussion and conclusions}

As we have shown above, the as-grown epitaxial \cobaltite\ films are
characterized by a certain amount of disorder and by a surface
atomic configuration that yields an oblique LEED pattern, consistent
with a preferential termination of the \cobaltite(110) surface in a
B plane. On the other hand, annealing improves the bulk and
interface crystalline order, smoothens the surface, and changes the
termination to a \cobaltite(110) A plane.  The transmission-like
spots in the RHEED patterns obtained during growth indicate that the
film roughens as it grows.  Such roughening can be a result of
kinetic limitations, for example a barrier for adatoms to descend
steps, or thermodynamics.  Since the lattice match is essentially
perfect, strain can be eliminated as a potential thermodynamic
driving force for roughening, leaving surface energetics as the only
viable driving force for roughening.  Because the \cobaltite\ (110)
surface is polar, the electrostatic contribution to its surface
energy increases linearly with film thickness. Therefore, one can
envision a growth sequence in which the film is initially flat
because the electrostatic contribution to the surface energy for a
film only a few layers thick is modest; as the film thickens,
however, the surface energy increases until the energy is reduced by
film faceting.  The problem with this explanation is that the
surface diffraction data indicates that the film still exposes
predominantly polar (110) facets.  The results, therefore, suggest
that the roughening is more the result of kinetic limitations during
growth than thermodynamics. The observation of the A termination of
the substrate and the film after annealing, but the B termination
during growth, suggests a potential kinetic limitation to forming
the apparently lower energy A termination during growth.

For the B termination, where the background in the LEED patterns is
high, it is possible that subsurface layers may contribute to the
LEED pattern, and the broad diffraction spots, also observed in the
RHEED data, suggest disorder or small domains for this surface. For
the annealed surface (A termination), however, the LEED background
intensity is very low, with very sharp diffraction spots. The
electron mean free path is at a minimum at around 50 eV in LEED, and
our data shows that in the range from 40 eV to 150 eV there are no
changes in the LEED pattern and no discernible increase in the
background intensity, indicating that we are probing the topmost
surface layer and that the presence of a disordered surface layer
can be ruled out.

Although the \spinel (110) substrate, the as-grown B terminated
\cobaltite, and the annealed A terminated \cobaltite\ are all polar
surfaces, only ($1\times 1$) surface diffraction patterns were
observed. The divergent electrostatic surface energy of polar
surfaces can be compensated in a variety of ways, including surface
reconstructions, surface defects, and adsorption of polar molecules
at the surface \cite{Noguera00,GFN08}. Several theoretical
calculations of the surface energies of the (110) surface of
\spinel\ consider significant modifications of the bulk termination
surface to remove the charge dipole, for instance by removing half
the atoms in the type-A termination (Fig.~\ref{fig:spinel_surface})
or considering a surface composed of O only (which yields the lowest
surface energy) \cite{DPW94,Harding99,FPW00,GFN08}. These
calculations also predict large surface relaxations of the surface
atoms; in addition, surface inversion in the spinel structure,
whereby B$^{3+}$ ions swap position with A$^{2+}$ ions of the layer
below, also results in significantly lower surface energies
\cite{DPW94}. The observed ($1\times 1$) LEED patterns rule out any
compensation mechanism that involves a periodic surface
reconstruction. While these patterns are still compatible with a
modified surface composition which keeps the surface unit cell
periodicity, this must be achieved while achieving charge
compensation. This issue has been addressed in the literature for
MgAl$_2$O$_4$(110) surfaces \cite{DPW94,Harding99,FPW00,GFN08},
which show that charge compensation leads to large modifications to
the surface composition; this would lead to larger surface unit
cells, and to smaller unit cells in LEED, which is not observed. The
adsorption of molecules at the surface should also be considered as
a possible mechanism leading to stabilization of the A and B
terminations. For example, it is now recognized that the long-range
$1\times 1$ ordering of the polar (000$\bar{1}$) surface of ZnO is
likely due to hydroxylation of the surface
\cite{KBW03,KGBW03,Woll04,Woll07}. In the present case, adsorption
of H in the form of protons could screen the charge on the
B-terminated surface, which would be negatively charged for the bulk
termination, while water dissociation could result in OH$^-$
radicals and screen the A-terminated surface, which is positively
charged for the bulk truncation. While this could explain the
shoulder in the 1s O peak of the \cobaltite\ XPS spectra
\cite{HU77,PMCL08}, it is unlikely that they would give the same
signature in the two cases. In fact, the shoulder is identical for
both terminations, with little variation in energy and intensity for
both the as-grown samples, measured immediately after growth, and
for the annealed samples, which are cleaned in an O-plasma after
been exposed to air (this shoulder is often assigned to adsorbed O
\cite{CBR76,JD79,KBR+84}). Therefore, assignment of the shoulder to
H on the surface cannot provide the needed screening for both
surfaces. Instead, we propose that the surfaces are compensated by
an ionic exchange mechanism, which in the case of \cobaltite\
reduces to filling and depleting surface electronic states on
opposite sides of the film. An estimate of the degree of charge
transfer can be obtained by assuming the formal charges on the ions.
As described in the Introduction, along the [110] direction the
spinel structure is composed of alternating, equally spaced A$^{2+}$
and B$^{2-}$ planes.  In this case, the thickness dependent
component of the surface dipole can be eliminated by placing charges
of $\pm$1 per surface unit cell on opposite sides of the crystal
\cite{Noguera00}. For the spinel, changing the stoichiometry of the
two outermost layers to MgAl$_5$O$_8$ and Mg$_3$Al$_3$O$_8$ can
provide the necessary $\pm$1 charges. For \cobaltite\ this same
charge compensation is achieved by transferring an electron from one
side of the crystal to the other, thereby oxidizing one Co$^{2+}$ to
Co$^{3+}$ on one side, and reducing one Co$^{3+}$ to Co$^{2+}$ on
the opposite side; or in a more general picture, filling a surface
state on one side and depleting it on the other.  As noted by
Noguera \cite{Noguera00}, more precise modeling of the charges on
the ions does not appreciably change this picture. Because the
electron scattering cross-sections are nearly the same for Al and
Mg, and are unchanged for \cobaltite, this model is consistent with
the observed ($1\times 1$) diffraction patterns.

The model proposed above suggests that the A-terminated annealed
surface should have a unique filled surface state that does not
exist in the bulk, and may confer the surface with unique electronic
and magnetic properties in a manner analogous to the metallic
behavior observed for LaAlO$_3$/SrTiO$_3$ interfaces.  We are
currently further investigating this possibility using ultraviolet
photoemission spectroscopy. Real space images of the atomic surface
order, for instance as imaged by AFM or STM, will also be
particularly useful in developing a more complete picture of the
charge compensation mechanisms at work.

In summary, we have demonstrated the growth of high quality
epitaxial \cobaltite(110) thin films on \spinel(110) by oxygen
assisted molecular beam epitaxy. Post-annealing is found to improve
considerably the film characteristics, as shown by several
reciprocal and real space probing techniques. Two different surface
terminations of the \cobaltite(110) surface are observed for the
as-grown and annealed films, which are associated with the two
possible terminations of the bulk \cobaltite(110).

\section*{Acknowledgements}

The authors acknowledge financial support by the NSF through MRSEC
DMR 0520495 (CRISP), MRSEC DMR 0705799, the ONR (C.H.A.), the
Petroleum Research Foundation Grant Numbers 42178-AC5 (J.W. and
E.I.A.) and 42259-AC5 (B.J.A. and U.D.S.), the DOE Catalysis and
Chemical Transformations Program, Grant Number DOE DE-FG02-06ER15834
(M.Z.B., T.S., E.I.A. and U.D.S.).

\bibliographystyle{elsart-num}
%\bibliography{c:/users/cav25/latex/general}



\end{document}